# Magic angle in thermal conductivity of twisted bilayer graphene

Yajuan Cheng[1], Zheyong Fan[2], Tao Zhang[1], Masahiro Nomura[3], Sebastian Volz[3], Guimei Zhu[4,*], Baowen Li[5,6,*], Shiyun Xiong[7,*]

1, School of Physics and Materials Science, Guangzhou University, Guangzhou 510006, China
2, College of Physical Science and Technology, Bohai University, Jinzhou 121013, China
3, Laboratory for Integrated Micro Mechatronic Systems (LIMMS/CNRS-IIS), The University of Tokyo, Tokyo 153-8505, Japan
4, School of Microelectronics, Southern University of Science and Technology, Shenzhen 518055, China
5, Department of Materials Science and Engineering, Department of Physics, Southern University of Science and Technology, Shenzhen 518055, China
6, Paul M. Rady Department of Mechanical Engineering and Department of Physics, University of Colorado, Boulder, Colorado 80305-0427, USA
7, Guangzhou Key Laboratory of Low-Dimensional Materials and Energy Storage Devices, School of Materials and Energy, Guangdong University of Technology, Guangzhou 510006, China
*Email: zhugm@mail.sustech.edu.cn; libw@sustech.edu.cn; xiongshiyun216@163.com

**Abstract:** We report a local minimum in thermal conductivity in twisted bilayer graphene (TBG) at the angle of 1.08°, which corresponds to the 'magic angle' in the transition of several other reported properties. Within the supercell of a moiré lattice, different stacking modes generate phonon scattering sites which reduce the thermal conductivity of TBG. The thermal magic angle arises from the competition between the delocalization of atomic vibrational amplitudes and stresses on one hand, and the increased AA stacking density on the other hand. The former effect weakens the scattering strength of a single scatterer while the latter one increases the density of scatterers. The combination of these two effects eventually leads to the apparition of the highlighted irregularity in heat conduction. The manifestation of a magic angle, disclosing new thermal mechanisms at nanoscale, further uncovers the unique physics of two-dimensional materials.

## 1. Introduction

Twisted bilayer graphene (TBG) exhibits a moiré pattern with a larger second lattice periodicity. When the twist angle between the two graphene layers reaches 1.08°, band hybridization and avoided crossings emerge and result in the formation of flat bands near the Dirac point[1-3]. This unusual behavior termed as 'magic angle', leads to novel phenomena that are not prevalent either in a single-layer or in a bilayer graphene. Among others are electronic correlation, superconductivity, spontaneous ferromagnetism, quantized anomalous Hall states, and topologically protected states. This magic angle has attracted substantial research interests in recent years since its theoretical prediction and experimental observation[4-13].

Up to now, most of the studies on TBG are focused on the electronic properties and less attention has been paid to thermal transport properties[14-18]. Considering that the single-layer graphene possesses excellent thermal conductivity (TC) ~3000-5000 W/mK at room temperature[19-21] and has broad applications in thermal management, it is also important and interesting to clear how the thermal transport properties depend on the twist angle. Since the twist of bilayer graphene can generate a second period, which is similar to phononic crystals, the thermal transport properties of TBG should be twist angle-dependent. In fact, new hybrid folded phonons will be generated in TBG due to the mixing of phonon modes from different high-symmetry directions of the two layers[22, 23]. As a result, the phonon properties are likely to be affected by the twists, especially for the out-of-plane modes, which is the main heat carrier in graphene[24].

In this work, we report an abnormal behavior in the vicinity of 1.08°, where the TC displays a local minimum. Our systematic investigations with homogeneous non-equilibrium molecular dynamic (HNEMD) simulations reveal that the TC minimum arises from the competition between the spatial dependence of atomic vibrational amplitude and stresses on one side, and the density of scattering sites on the other side. The former changes the scattering strength of a single scattering site while the latter alters the number of scattering sites.

## 2. Structure and Simulation Methods

The lattice parameter of the Moiré lattice in TBG that forms due to the twisting of two graphene layers, decreases with increasing twist angle. The unit cell of the Moiré lattice takes a hexagonal shape as shown by the black lines in the top panel of Fig. 1. The relation between the Moiré lattice parameter $a_{moire}$ and the twist angle $\theta$ can be expressed as $a_{moire} = \frac{a}{2\sin\left(\frac{\theta}{2}\right)}$, $a = 2.452$ Å being the lattice parameter of monolayer graphene. Inside the unit cell of the Moiré lattice, the stacking between the two layers is not uniform. Typical stacking modes include the AA, AB, and SP stacks, which are schematically illustrated in the bottom panel of Fig. 1. An orthogonal simulation box was constructed based on a larger unit cell with a rectangle shape as indicated by the blue dashed lines in Fig. 1.

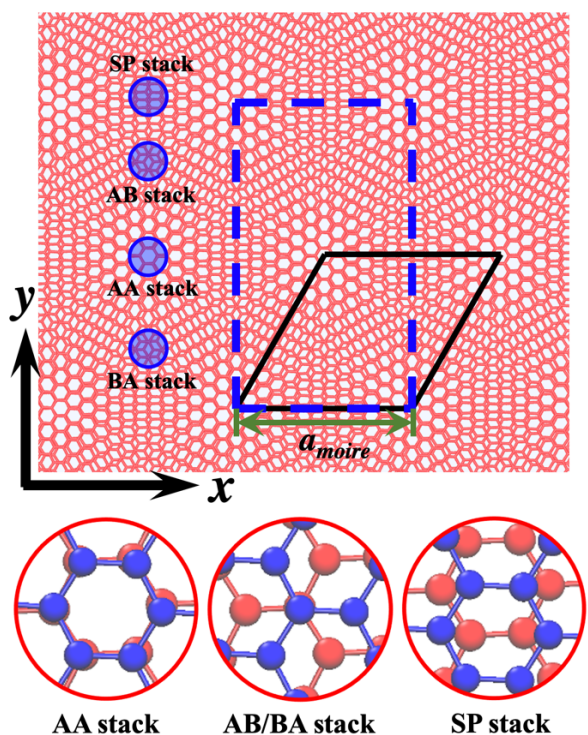

**Figure 1**. Top panel: the Moiré lattice formed in TBG. The relative position of AA, AB, and SP stacks are illustrated by the circles and the unit cell of the Moiré lattice is indicated by the black parallelogram. Bottom panel: the atomic arrangements of AA, AB, and SP stacks. Red and blue atoms correspond to the atoms in the bottom and top layers, respectively.

To simulate the thermal transport properties of TBG, we performed homogeneous nonequilibrium molecular dynamics (HNEMD)[25, 26] simulations implemented in the graphics processing units molecular dynamics (GPUMD) package[27]. The intralayer C-C interaction is described by the optimized Tersoff potential[28] while the interlayer C-C interaction is governed by the interlayer potential (ILP)[29]. The ILP potential has been proven to provide a better description for the interlayer weak Van der Waals interactions than the commonly used Lennard-Jones (LJ) potential as demonstrated by Lebedeva *et al* [29]. The cutoff of the ILP potential is chosen as 15 Å, beyond which atomic interaction energies are negligible. During the simulations, a time step of 1.0 fs is adopted for the integration of equations of motion. Due to the symmetry of graphene, the structure of TBG with a twist angle $\theta$ larger than 30° is identical to the structure with the angle 60° - $\theta$. As a result, we only consider the twist angles from 0° to 30°. The twist angle of 0° corresponds to the AB stack, which is the most stable phase of bilayer graphene. Due to the different unit cell size at different twist angles, the size of the simulation box is twist-angle dependent. The detailed size for each twist angle can be read from the *x* and *y* dimensions in Fig. S4.

In HNEMD simulations, the driving force is chosen as $F_e = 1.0 \times 10^{-5}$ Å$^{-1}$ for all twist angles. The running TC as a function of the production time *t* is evaluated as:[25]

$$\kappa(t) = \frac{1}{t}\int_0^t d\tau \frac{\langle J_q(\tau)\rangle_{ne}}{TVF_e} \quad (1)$$

where *V* and *T* refer to the system volume and temperature, respectively. $F_e$ is the driving force, which leads to the generation of nonequilibrium heat current $\langle J_q(\tau)\rangle_{ne}$ in the system. Since the Moiré lattice possesses C$_6$ symmetry, the in-plane thermal transport is isotropic. Consequently, we computed the TC along both the *x* and *y* directions (the basal plane) with 10 ns of simulations for each direction. With two independent simulations for each sample, the final TC is averaged by 4 data sets.

For two-dimensional materials, the out-of-plane vibrations can contribute dramatically to the total TC. To demonstrate how the twist angle affects the TC contributed by in-plane and out-of-plane modes, it is necessary to decompose the total TC into the in-plane and out-of-plane modal contributions. To this end, we decompose the heat current into the in-plane and out-of-plane components:[25]

$$\boldsymbol{J} = \boldsymbol{J}^{in} + \boldsymbol{J}^{out} \quad (2)$$

where

$$J^{in} = \sum_i \sum_{j \neq i} r_{ij} \left( \frac{\partial U_j}{\partial x_{ji}} v_{xi} + \frac{\partial U_j}{\partial y_{ji}} v_{yi} \right) \qquad (3)$$

and

$$J^{out} = \sum_i \sum_{j \neq i} r_{ij} \left( \frac{\partial U_j}{\partial z_{ji}} v_{zi} \right) \qquad (3)$$

$U_j$ refers to the potential energy of atom $j$ and $r_{ij}$ corresponds to the vector between atoms $i$ and $j$. $v_{\alpha i}$ and $\alpha_{ji}$ ($\alpha = x, y, z$) denote the velocity of atom $i$ along the direction $\alpha$ and the projected distance along the $\alpha$ direction between atoms $i$ and $j$, respectively.

## 3. Results and discussions

Fig. 2(a) illustrates the twisted angle-dependent TC at 300 K. In general, the TC of TBG is reduced compared to the untwisted one. Below 1.08°, the TC decreases rapidly with the increase of the twist angle. In contrast, beyond 10°, the TC slightly increases, which is in agreement with other simulations[16-18] and experiments[14, 15]. Surprisingly, we observe an abnormal TC valley in the vicinity of 1.08°, beyond which the TC rapidly increases and reaches a local maximum when the angle reaches around 3°. At angles larger than 3°, the TC decreases to a plateau.

It is well known that 1.08° corresponds to the magic angle for electron transport, at which superconductivity appears at low temperatures[2, 3]. Thus we also label this angle as 'thermal magic angle'. We now investigate the operating physical mechanisms at the origin of this singularity.

We first acknowledge that the thermal magic angle also emerges at other temperatures as shown in Figs. 2(c) and (d). Moreover, we also confirm that the effect is not specific to the adopted interlayer potential by calculating the TC of TBG with the Lennard-Jones (LJ) potential. The results emphasize that an abnormal TC increase starting from 1.56° is also preserved (Fig. S2). Those preliminary outcomes confirm that the observed TC minimum can be considered as an intrinsic property of TBG.

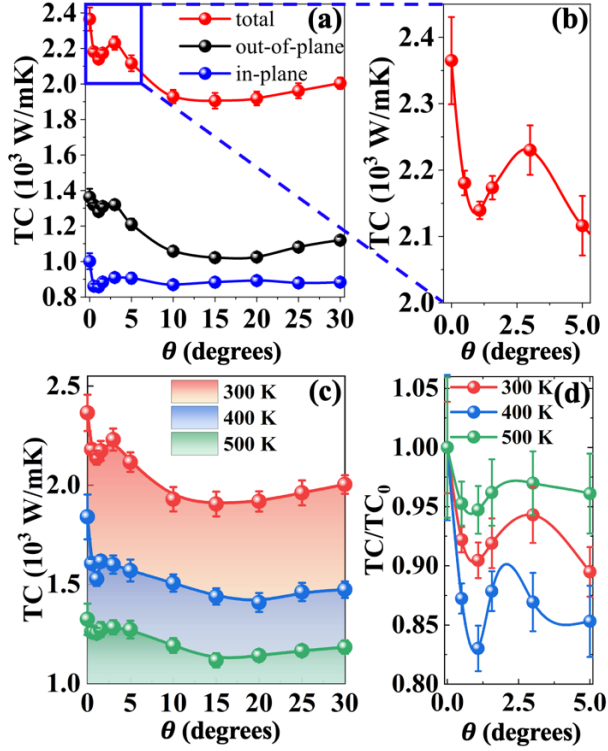

**Figure 2**. (a) Total, in-plane, and out-of-plane thermal conductivity of TBG as a function of twist angle from 0° to 30° at 300 K. (b) Total thermal conductivity of TBG versus twist angle below 5°. (c) Total thermal conductivity of TBG versus with twist angle at temperatures 300 K, 400 K, and 500 K. (d) Normalized thermal conductivity with respect to the value of the untwisted structure as a function of twist angle at 300 K, 400 K, and 500 K.

In two-dimensional materials, the in-plane and out-of-plane modes contribute much differently to the total TC[30]. To examine the effect of the twist angle on the in-plane and out-of-plane modes, we decompose the total TC into both contributions at 300 K following the method proposed in Ref.[25] and yielding the results of Fig. 2(a). Similarly to other two-dimensional materials, the out-of-plane modes contribute more to the total TC than the in-plane ones. Interestingly, the in-plane TC is more sensitive to the twist angle below 1.08° while remaining almost constant beyond 3°. As a result, the in-plane TC around 1.08° is largely contributing to the TC minimum. In contrast, the out-of-plane TC rapidly decreases from 3° to 10° and increases beyond 15°, governing the overall TC in this angle range.

In TBG, the Moiré unit cell is large and the stacking is not uniform inside the unit cell. It changes from the stable AB stack to the unstable AA and SP stacks continuously (Fig. 1). Such a non-uniform stack strongly scatters phonons, thus reducing the TC. On the other hand, the stack also leads to a specific spatial distribution of vibrational and mechanical properties, such as the atomic vibrational amplitude and stress. We now show that the thermal magic angle arises from the interplay between the increased number of the scattering centers on one side, and the delocalization of the vibrational amplitude and stress on the other side.

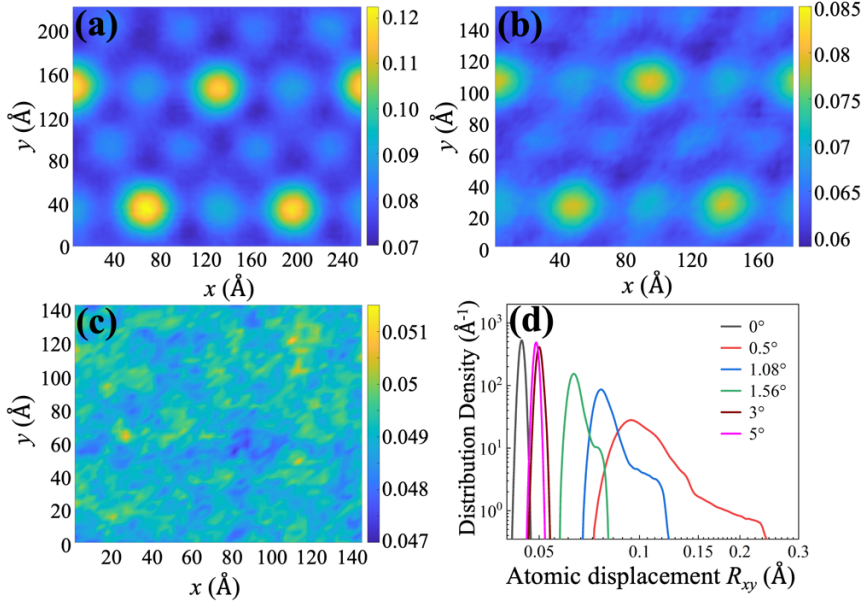

**Figure 3**. Spatial distribution of the in-plan atomic vibrational amplitude at (a) $\theta =1.08°$, (b) $\theta =1.56°$, and (c) $\theta =5.0°$ (unit: Å). The bright yellow regions correspond to AA stacks in (a) and (b). (d) the atomic in-plane vibrational amplitude distribution at different angles.

At zero temperature, the AA stack has a larger equilibrium interlayer distance (3.65 Å) compared to the one of the AB stack (3.38 Å). At small twist angles, the AA and AB stack centers are significantly remote, and the interlayer distances at AA and AB centers highlight differences (Fig. S4). With the increase of $\theta$, the distance between AA and AB centers reduces rapidly. To maintain the same interlayer distances at the AA and AB centers in the same case of small angles, a larger stress is required. The increased stress eventually makes the difference between the interlayer distance in the vicinity of the AA and AB regions smaller, i.e., the distribution of interlayer distances becomes narrowed. Since the interlayer interaction strength between two graphene layers depends on the interlayer distance, and the atomic vibration is directly related to the atomic interaction strength, the non-uniform interlayer distance yields an also non-uniform vibrational amplitude distribution. To examine those vibrations, we analyze the atomic trajectories and extract the time-averaged atomic vibrational amplitude for each case. The instantaneous vibrational amplitude of each atom was calculated as the distance between its instantaneous and equilibrium positions. Due to the fluctuations of the TBG position in the out-of-plane direction ($z$), we only count the averaged atomic vibrational amplitude along the in-plane directions ($x$ and $y$), which is noted as $R_{xy}$.

Figs. 3(a)-(c) report the time-averaged atomic in-plane vibrational amplitude $R_{xy}$ at 1.08°, 1.56° and 5° (300 K). At small angles, the spatial distribution of $R_{xy}$ clearly represents the Moiré patterns, where $R_{xy}$ is maximized at the AA stack regions. The mismatch in the vibrational amplitudes in different regions reduces rapidly as the twist angle increases. In Fig. 3(d), the distribution range of $R_{xy}$ is large at small angles but reduces rapidly with the increase of the twist angle. Starting from 3°, $R_{xy}$ becomes a normal distribution with a small standard deviation, which is similar to the one of the untwisted structure. The averaged $R_{xy}$ at $\theta > 3°$ reaches 0.05Å, which is slightly larger than the corresponding value of the untwisted structure (Fig. S7).

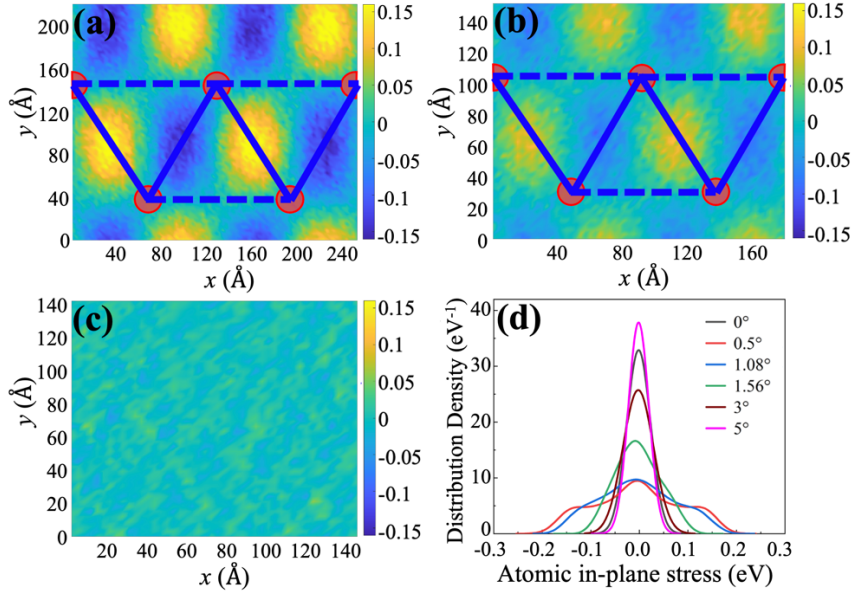

**Figure 4**. Spatial distribution of the atomic (normal) stress along the $x$ direction ($\sigma_{xx}$) of the bottom graphene layer for (a) $\theta$ =1.08°, (b) $\theta$ =1.56°, and (c) $\theta$ =5° at 300 K (unit: eV). The red disks denote the AA stack centers. (d) the distribution of atomic normal stress for all atoms at indicated angles. The averaged stress for all systems is shifted to zero.

Similar to $R_{xy}$, we find that the in-plane atomic stress is also space dependent because the rotation changes the periodicity of graphene along a specific direction, leading to different period lengths along the same direction for the top and bottom layers. The lattice mismatch can cause the atoms from the top and bottom layers to attract or repel each other along a specific in-plane direction. Since the attraction and repulsion occur between the two layers, the stresses of the top and bottom layers are directed in opposite directions with similar magnitudes at the same position.

Figs. 4(a)-(c) illustrate the time-averaged normal stress along the $x$ direction, i.e., $\sigma_{xx}$, of the bottom layer graphene at 300 K for $\theta$ = 1.08°, 1.56° and 5°, respectively. In contrast to the distribution of maximum/minimum of vibrational amplitudes that are located at the AA or the AB stack, the maximum and minimum of the stresses are located at the SP stack regions, i.e., the middle of two neighboring AA regions.

Due to the symmetry of the TBG, the maximum/minimum of the normal stresses $\sigma_{xx}$ and $\sigma_{yy}$ are located at the two edges of the triangle formed by AA centers (solid lines in Fig. 4(a)), while another edge of the triangle (dashed line) possesses the maximum/minimum shear stress $\tau_{xy}$ as shown in Fig. S11. At small twist angles, heterogeneity in atomic stress distribution is observed. Such local stress differences reduce rapidly with the increase of the twist angle and become negligible eventually.

To check the distribution of atomic stresses, we counted the distributions of $\sigma_{xx}$ and $\sigma_{yy}$ as shown in Fig. 4(d). It clearly shows that the stress is distributed over a broad range in the small twist angle region. For example, $\sigma_{xx}$ and $\sigma_{yy}$ are ranging from -0.2 to 0.2 eV when $\theta$ = 0.5°. With the increase of $\theta$, the stress distribution range reduces and it becomes a normal distribution with reduced standard deviations starting from 5°, which is similar to the value of the untwisted structure (Fig. S10).

The structure-induced spatial inhomogeneity in TBG can strongly scatter phonons, which in turn reduces the TC in bilayer graphene with a twist. Larger vibrational amplitudes indeed induce the exploration of the asymmetric or the anharmonic range of the interatomic potential energy. As a result, the anharmonicity of the system augments, as is the case at high temperatures[31]. On the other hand, the existence of stress will shift phonon frequencies due to the modification of force constants. As a result, local stresses can lead to the mismatch of phonon frequencies in different regions[32, 33]. The total phonon scattering strength depends both on the scattering strength of each individual scatterer and of the density of scatters. The larger property difference at the AA and AB stacks, the stronger of scattering effect by the AA or AB centers. Since the large property difference at the AA and AB sites will lead to broad distribution of the corresponding property, the scattering strength of a single scatterer can be characterized by the standard deviation of the corresponding distribution function in TBG, namely $std(R_{xy})$ and $std(\sigma)$. On the other hand, since the number of stress and vibrational amplitude maxima/minima is directly related to the AA stack numbers, we can use the AA stack density $\rho_{AA}$ to represent the density of scatterers. Based on such a consideration, we now define a *S*-factor:

$$S = \frac{C}{std(R_{xy})std(\sigma)\rho_{AA}} \quad (4)$$

to quantify the effect of the scattering, where *C* is a constant. The denominator $std(R_{xy})std(\sigma)\rho_{AA}$ characterizes the total scattering strength, i.e., the product of scattering strength of a single scatterer with the site density of scatterers.

The variation of $\rho_{AA}$, $std(R_{xy})$ and $std(\sigma)$ as a function of $\theta$ are illustrated in Fig. 5. At small angles, $\rho_{AA}$ increases dramatically with $\theta$ while both $std(R_{xy})$ and $std(\sigma)$ decrease rapidly. The increased $\rho_{AA}$ produces more scattering sites while the reduced $std(R_{xy})$ and $std(\sigma)$ weakens the scattering strength of a single scattering site. As a result, the two effects compete with each other and eventually lead to the abnormal TC increase beyond 1.08°.

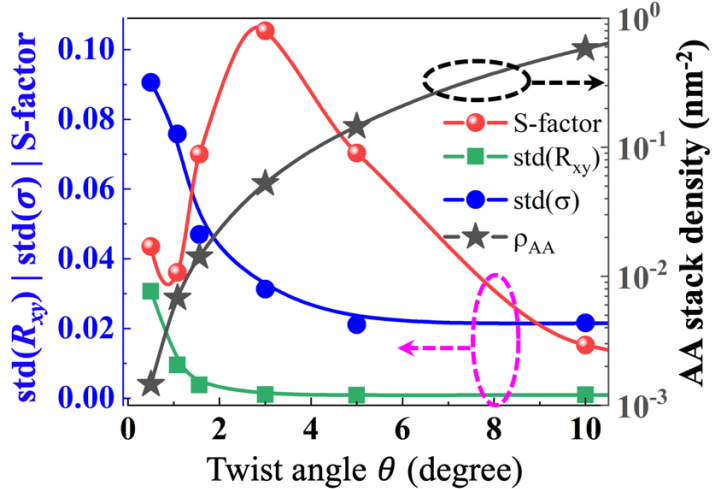

**Figure 5**. Twist angle-dependent S-factor in TBG calculated by Eq. 4. The constant *C* is chosen as $6 \times 10^{-6}$ ÅeV/nm² to match the value of the *S*-factor in the range of the $std(\sigma)$.

Fig. 5 also illustrates the variation of the *S*-factor with the twist angle at 300 K. Interestingly, the twist angle-dependent *S*-factor features the same trend as the TC. the *S*-factor decreases first from 0.5° and reaches a local minimum around 1.08°. With the further increase of the twist angle, the *S*-factor increases and reaches a local maximum around 3°. Then, it reduces with the increase of the twist angle. The similar trend and the identical transition angles between the *S*-factor and the TC confirms that the TC minimum arises from the interplay between the reduced scattering strength of a single scatterer and the increased scattering site density.

With the further increase of the twist angle after 3°, the distributions of both the vibrational amplitude and stress only change slightly with twist angle (Figs. 3(d) and 4(d)), indicating that the weakening of the scattering strength is negligible with the further increase of the twist angle. As a result, the TC beyond 3° is governed by the scattering site density increase again, which finally leads to the diminution of TC after 3°. We note that beyond 10°, the TC is slightly enhanced with the increase of the twist angle, which is presumably related to a coherent phonon transport effect similar to the case of superlattices[34-36].

Beyond 10°, all properties become delocalized (space independent). The distance between AA centers shrinks to sizes down to ~1.5 nm at 10°, which is likely to be shorter than the size of phonon wave packets. In such a situation, those phonons do not interact with the scatterers and behave as if the material was homogeneous. With the increase of $\theta$, more phonons travel coherently, which leads to the increase of TC[34, 36].

It is worth noting that although both electronic and thermal magic angles take the same value of 1.08°, the underlying mechanisms are clearly different. In the electronic case, the involved strong electron correlation happens only at extremely low temperature, at which lattice vibrations are extremely small and will not break down those correlations. In contrast, in the thermal case, atomic vibrations are driving the TC minimum which even exist at high temperatures as illustrated in our Fig. 2.

## 4. Conclusions

In summary, we uncover a thermal magic angle of 1.08° at which a local minimum of the TC appears. The decomposition of the TC demonstrates that its rapid reduction below 1.08° arises from the diminution of both in-plane and out-of-plane mode contributions. A TC reduction arises mainly from the out-of-plane modes beyond 3°. The twist in bilayer graphene leads to non-uniform stackings, resulting in the localization of atomic vibrations which consequently scatters phonons. At small angles, both vibrational amplitude and stress are delocalized with the increase of the twist angle, leading to a reduced scattering strength of a single scatterer. On the other hand, the number of scattering sites dramatically increases with the twist angle. The competition between those two effects eventually results in the formation of a thermal magic angle. The highlighted physical mechanisms will provide new degrees of freedom in thermal management and control by using graphene and are more generally uncovering unexpected and transversal behaviors in two-dimensional materials.

## 5. Acknowledgements

Y. Cheng and Z. Fan contribute equally to this work. This work was supported by the National Natural Science Foundation of China under Grant No. 12174276 (SX) and No.52250191 (BL), the Major Research Plan of the National Natural Science Foundation of China (Grant No. 91833303), and the Major International (Regional) Joint Research Project of the National Natural Science Foundation of China (Grant No. 51920105005).

## 6. Conflicts of interest

The authors declare no conflicts of interest.